\documentclass[prd,aps,showpacs,preprintnumbers,amssymb]{revtex4}
\usepackage{axodraw}
\usepackage{color}
\usepackage{epsf}

\def\e3p{$\eta \rightarrow 3 \pi$}

\begin{document}
\title{%
\hfill{\normalsize\vbox{%
\hbox{}
 }}\\
{  The abelian Higgs model and its phase transitions revisited}}

\author{Renata Jora
$^{\it \bf a}$~\footnote[2]{Email:
 rjora@theory.nipne.ro}}

\affiliation{$^{\bf \it a}$ National Institute of Physics and Nuclear Engineering PO Box MG-6, Bucharest-Magurele, Romania}

\date{\today}

\begin{abstract}
The  abelian Higgs model and its phase structure are discussed from the perspective  that the gauge and scalar fields admit a dual description in terms of fermion variables. The results which indicate the presence of three main phases: Coulomb, Higgs and confinement agree well with those in the literature although a nonstandard order parameter is employed.
\end{abstract}
\maketitle

\section{Introduction}

The $U(1)$ abelian Higgs model is one of the simplest theories that contains both gauge fields and matter. The phase structure for this model has been studied at zero temperature both for compact Higgs $|H|={\rm const}$ \cite{Fradkin}-\cite{King} or for varying $|H|$ \cite{Gerdt}. The finite temperature regime has also been analyzed \cite{Banks},\cite{Susskind}.  It was shown in all these instances that the model displays three main phases: a Coulomb phase where the potential is $V(r)\approx \frac{1}{r}$;  a Higgs phase where $V(r)\approx {\rm const}$ and a confinement phase where $V(r)\approx r$. However for the particular case of a Higgs in the fundamental representation with charge unit it seems that there is no real distinction between the Higgs phase and the confinement one and thus the two of them are connected. All the above findings have been confirmed numerically through Monte Carlo simulations \cite{Carson}.

In the present work we shall revisit the abelian Higgs model without making any constraining assumptions by considering its dual description in terms of fermion variables. We suggest that this alternate description may alter the structure of the Lagrangian expressed in terms of the original variables.  The relevant parameters  are the coefficients of the terms in the modified Lagrangian and their relative magnitudes reveals the exact phase in which the system is in.  We then use the standard path integral approach to show that the Lorentz invariant function
$\langle A^{\mu}(x)A_{\mu}(y)\rangle$ (but our arguments would work as well for the regular two point function) is a reasonable order parameter to indicate the behavior of the model in different phases.
Although we use a novel perspective our findings agree well with the standard knowledge in the field.

\section{The set-up}

We start with the $U(1)$ abelian Higgs model given by the Lagrangian:
\begin{eqnarray}
{\cal L}=-\frac{1}{4}F_{\mu\nu}F^{\mu\nu}+D^{\mu}\Phi^*D_{\mu}\Phi-V(\Phi),
\label{lagr456}
\end{eqnarray}
where,
\begin{eqnarray}
&&D_{\mu}\Phi=(\partial_{\mu}+ie A_{\mu})\Phi
\nonumber\\
&&V(\Phi)=m^2\Phi^*\Phi+\frac{\lambda}{2}(\Phi^*\Phi)^2.
\label{expr6578}
\end{eqnarray}

Upon expansion of the gauge kinetic term for the scalar field the Lagrangian in Eq. (\ref{lagr456}) becomes:
\begin{eqnarray}
&&{\cal L}=-\frac{1}{4}F_{\mu\nu}F^{\mu\nu}+\partial^{\mu}\Phi^*\partial_{\mu}\Phi-i e A^{\mu}\Phi^*\partial_{\mu}\Phi+ieA_{\mu}\Phi\partial^{\mu}\Phi^*+{\cal L}_1
\nonumber\\
&&{\cal L}_1=e^2A_{\mu}A^{\mu}\Phi^*\Phi-m^2\Phi^*\Phi-\frac{\lambda}{2}(\Phi^*\Phi)^2.
\label{rez43567}
\end{eqnarray}

Now consider the dual description of the scalar and gauge fields in terms of fermion variables:
\begin{eqnarray}
&&A_{\mu}=\frac{1}{M^2}\bar{\Psi}\gamma_{\mu}\Psi
\nonumber\\
&&\Phi=\frac{1}{M^2}(\bar{\Psi}\Psi+\bar{\Psi}\gamma^5\Psi)
\label{rez4332456}
\end{eqnarray}
Note that we can do this and still preserve the gauge invariance. Moreover the corresponding bilinear forms can be considered independent since an on-shell fermions has four degrees of freedom, an on shell massless gauge boson has two and the complex scalar has two. This means that the redefinition of the fields made in Eq. (\ref{rez4332456}) matches the number of original degrees of freedom for the Lagrangian in Eq. (\ref{lagr456}).

We shall work with the Lagrangian ${\cal L}$ in which we ignore the $\lambda$ interaction term  the scalar mass $m^2$ is an infinitesimal parameter. We consider  the Fierz transformation,
\begin{eqnarray}
&&\Phi\Phi^*=\bar{\Psi}\Psi\bar{\Psi}\Psi-\bar{\Psi}\gamma^5\Psi\bar{\Psi}\gamma^5\Psi=
\nonumber\\
&&=\frac{1}{2}[\bar{\Psi}\gamma^{\mu}\Psi\bar{\Psi}\gamma_{\mu}\Psi-\bar{\Psi}\gamma^{\mu}\gamma^5\Psi\bar{\Psi}\gamma_{\mu}\gamma^5\Psi]=\frac{1}{2}[A^{\mu}A_{\mu}-A^{\mu 5}A^5_{\mu}]
\label{rez2121}
\end{eqnarray}
to write:
\begin{eqnarray}
&&A_{\mu}^2=xA_{\mu}^2+(1-x)[2|\Phi|^2+(A_{\mu}^{5})^2]
\nonumber\\
&&|\Phi|^2=y|\Phi|^2+(1-y)/2[A_{\mu}^2-(A_{\mu}^5)^2],
\label{res3221}
\end{eqnarray}
where $x$ and $y$ are real parameters  with $0\leq x \leq1$ and $0\leq y \leq1$.
Then the Lagrangian ${\cal L}_1$ will become:
\begin{eqnarray}
&&{\cal L}_1=e^2[x A_{\mu}^2+(1-x)[2|\Phi|^2+(A_{\mu}^{5})^2]][y|\Phi|^2+(1-y)/2[A_{\mu}^2-(A_{\mu}^5)^2]]-m^2|\Phi|^2=
\nonumber\\
&&=2e^2y(1-x)|\Phi|^4+e^2x(1-y)/2A_{\mu}A^{\mu}A_{\nu}A^{\nu}+e^2[x y+(1-x)(1-y)]A^{\mu}A_{\mu}|\Phi|^2-m^2|\Phi|^2.
\label{new6546}
\end{eqnarray}
Since the parameter $m^2$ is considered infinitesimal there is no need to make the replacement in Eq. (\ref{res3221}) also for the term $m^2|\Phi|^2$. We solve the equation of motion for the auxiliary field $A_{\mu}^5$ to get $A_{\mu}^5=0$. Thus in the last line of the equation we took $A_{\mu}^5=0$.
Net we shall analyze the phases of the Lagrangian described in Eq. (\ref{new6546}) in term of the parameters $x$ and $y$.

\section{The phase structure}

The phase diagram of the abelian Higgs model was discussed in \cite{Fradkin}-\cite{Carson} whereas the simple abelian gauge model was treated in \cite{Kogut2}, \cite{Banks2}.
In \cite{Fradkin} where a  lattice approach is considered the phases are determined by the magnitude of the parameters $\beta$ and $K$ where $K=\frac{1}{e^2}$ and $\beta=R^2$ where R is the Higgs length. Alternatively $K$ is the dimensionless parameter that multiplies the gauge kinetic term whereas $\beta$ is the parameter that multiplies the scalar gauge kinetic term.
It is then clear that for the Lagrangian given in Eq. (\ref{new6546}) the terms of interest are the $(A_{\mu}A^{\mu})^2$ and $|\Phi|^4$  which after the rescaling of the fields with the constants in front of them will  multiply the kinetic term for the scalar field with $1/[e^2y(1-x)]^{1/2}$ whereas that of the gauge field with $1/[e^2x(1-y)]^{1/2}$. Without loss of generality we shall consider the initial value of $e^2$ large such that by tuning $x$ and $y$ one can get any value of the effective coupling small or large.

Let us briefly explain our procedure. We denote $z_1=[x(1-y)]^{1/4}$ and $z_2=[y(1-x)]^{1/4}$.   We rescale the fields as $z_1A_{\mu}=A_{\mu}'$ and $z_2\Phi=\Phi'$ and write the Lagrangian in terms of the new variable in order to put in evidence the relative magnitude of the kinetic terms. We shall use an unusual order parameter (see \cite{King}for other options found in the literature) the Lorentz invariant two point function $\langle A^{\mu}(x)A_{\mu}(y)\rangle$ and work in the Feynman gauge. It is evident then that the order parameter is not gauge invariant. However we will show that the behavior of the model is very well indicated by this order parameter by using the standard functional approach.
In order to compute the order parameter we will go back to  the initial approximate Lagrangian where the variables $\Phi(x)$a nd $A_{\mu}(x)$ are retrieved.

In conclusion we find four limiting cases:

\subsection{Higgs phase}
This phase is obtained for the following values of the parameter $x$ and $y$:  $x\approx 0$ and $y \approx 0$ or $x\approx 1$ and $y \approx 1$.
In both these cases the full Lagrangian has the expression:
\begin{eqnarray}
&&{\cal L}=-\frac{1}{4}1/(z_1)^2F_{\mu\nu}'F^{\prime\mu\nu}+1/(z_2)^2\partial^{\mu}\Phi^{\prime*}\partial_{\mu}\Phi'-
\nonumber\\
&&-i e/(z_1z_2^2) A^{\prime\mu}\Phi^{\prime *}\partial_{\mu}\Phi'+ie/(z_1z_2^2)e A_{\mu}'\Phi'\partial^{\mu}\Phi^{\prime*}+e^2A_{\mu}'A^{\prime\mu}_1\Phi^{\prime*}\Phi'-m^2/z_2^2\Phi^{\prime*}\Phi'
\label{lagr434}
\end{eqnarray}
where $z_1\approx 0$ and $z_2\approx 0$. If we solve for the Higgs expectation value in the initial Lagrangian in Eq. (\ref{new6546}) we find:
\begin{eqnarray}
\langle \Phi^2\rangle=\frac{m^2}{4e^2y(1-x)}={\rm large}
\label{Higgs435}
\end{eqnarray}
Here it is considered that the limits of the parameters $x$ and $y$ supersede the limits small or large of the parameters $m^2$ and $e^2$. We this expect that this phase will correspond to the Higgs phase.

In order to show that indeed this situation corresponds to the Higgs phase we first rescale the Higgs field as $e\Phi\Rightarrow\Phi$. Then all the terms that contain the Higgs field in the Lagrangian in Eq. (\ref{lagr434}) will be further suppressed and can be neglected. Moreover we  shall use the Feynman gauge and calculate the two point function for the corresponding Lagrangian. First we need to compute the partition function in the Fourier space:
\begin{eqnarray}
Z\approx\int d A_{\mu}(p) d\Phi(q)\exp[i[-\frac{1}{2}\sum_pA^{\mu}(p)p^2A_{\mu}(-p)+\sum_{p,q,r}A^{\mu}(p)A_{\mu}(q)(\Phi_1(r)\Phi(-p-r-q)_1+\Phi(r)_2\Phi_2(-p-r-q)]],
\label{part456}
\end{eqnarray}
where $\Phi_1(x)={\rm Re}\Phi(x)$ and $\Phi_2(x)={\rm Im}\Phi(x)$. In order to find the two point function we shall use a trick. We change the variable $\Phi(p)=\lambda \Psi(p)$ for  any $p\neq q$ and $\Phi_2(p)=\lambda \Psi_2(p)$ where $q$ is arbitrary and fixed. Here the parameter $\lambda$ is considered very large. Then the partition function will become:
\begin{eqnarray}
&&Z={\rm const}\frac{1}{\lambda^N}\int d A_{\mu}(p) d\Phi(p)\exp[i[-\frac{1}{2}\sum_pA^{\mu}(p)p^2A_{\mu}(p)+\sum_pA^{\mu}(p)A_{\mu}(-p)\Phi_1(q)\Phi_1(-q)]]=
\nonumber\\
&&={\rm const}\frac{1}{\lambda^N}\int d A_{\mu}(r)\frac{1}{\sum_p A^{\mu}(p)A_{\mu}(-p)}\exp[i[-\frac{1}{2}\sum_pA^{\mu}(p)p^2A_{\mu}(p)]]
\label{rez324567}
\end{eqnarray}
where we took into account that $\Phi(-p)=\Phi(p)^*$.  Then using,
\begin{eqnarray}
\sum_{p^2}\frac{\delta}{\delta p^2}Z={\rm const}\frac{1}{\lambda^N}\int d A_{\mu}(p)\exp[i[-\frac{1}{2}\sum_pA^{\mu}(p)p^2A_{\mu}(p)]]={\rm const}\prod_{p}\frac{1}{(p^2)^2}
\label{rez11213}
\end{eqnarray}
we obtain:
\begin{eqnarray}
Z=a\prod_{p}\frac{1}{(p^2)^2}+b
\label{qw234}
\end{eqnarray}
where $a$ and $b$ are two constants independent on the momenta (For example $b$ takes into account the fact that $Z\neq 0$ even for $p^2=\infty$).
The Lorentz invariant two point function in the Fourier space can then be written as (the momentum delta function can be included from the beginning in the equation if we take into account the initial expression of the partition function in Eq. (\ref{rez324567})):
\begin{eqnarray}
\langle A^{\mu}(p)A_{\mu}(-p)\rangle=\frac{1}{Z}\frac{\delta Z}{\delta p^2}=-2a\frac{1}{(p^2)^3}\frac{1}{a\prod_{q}\frac{1}{(q^2)^2}+b}\prod_{q\neq p}\frac{1}{(q^2)^2}\approx {\rm const}\frac{1}{(p^2)^3}
\label{rez43567}
\end{eqnarray}
We need to justify the result in Eq. (\ref{rez43567}). For that we observe that whereas $a$ is finite the quantity $b$ measures the degree of divergence of the partition function so it is very large. Thus the $a$ term can be neglected compared to the $b$ one.

 We are mainly interested to find the potential corresponding to the order parameter in the coordinate space between two sources. In this approach we need to consider as sources two scalar fields with the momenta $q_1$, $q_2$ such that $p^2=(q_1-q_2)^2\approx |\vec{q_1}-\vec{q_2}|^2=|\vec{p}|^2$. Then we can write directly:
 \begin{eqnarray}
 V={\rm const} \frac{1}{b}\int_{-\infty}^{\infty} dq \frac{\exp[iqr]}{r} q\frac{1}{(q^2+\mu^2)^3}=\frac{{\rm const}}{r}\exp[-\mu r][r+O(\mu)]\approx {\rm  const}
\label{res876154}
\end{eqnarray}
Here the integral is done on a contour closed above in the complex plane with the calculation of the residue of the third order pole $q=i\mu$. The constant in front contains  a product of factors that goes to infinity or zero that lead overall to a finite constant. If the limit $\mu$ is taken to zero one regains the standard result for the HIggs phase which says that the potential is constant.

In the end  it is important to mention that   the gauge field acquires a mass in the Higgs phase as usual although this is not manifest in our approach because we integrated over the shifted scalar field which corresponds to the initial fields in the Lagrangian and not to that resulting from the spontaneous symmetry breaking.  Note also that in this approach the Higgs phase is present even in the absence of the $\lambda$ term in the Lagrangian.

\subsection{The Coulomb phase}
This phase is obtained for  $y \approx 1$ and $x \approx 0$.
The approximate Lagrangian is:
\begin{eqnarray}
&&{\cal L}=-\frac{1}{4}1/(z_1)^2F_{\mu\nu}'F^{\prime \mu\nu}+1/(z_2)^2\partial^{\mu}\Phi^{\prime*}\partial_{\mu}\Phi'-
\nonumber\\
&&-i e/(z_1z_2^2) A^{\prime\mu}\Phi^{\prime*}\partial_{\mu}\Phi'+ie/(z_1z_2^2)e A_{\mu}'\Phi'\partial^{\mu}\Phi^{\prime*},
\label{lagr434}
\end{eqnarray}
where $z_1\approx 0$ and $z_2=\approx 1$. There is no mass for the gauge boson and the vev of the Higgs is zero.  With the change of variable $e\Phi\rightarrow\Phi$ the scalar field decouples and the final Lagrangian contains in first order only the kinetic term for the gauge field. The propagator is simply $\frac{1}{p^2}$ and the potential in the coordinate space is $V(r)\approx \frac{1}{r}$. This is the Coulomb phase of the abelian Higgs model.

\subsection{ Higgs +confinement phase}
This case corresponds to  $y \approx 0$ and $x \approx 1$ and  has the Lagrangian:
\begin{eqnarray}
&&{\cal L}=-\frac{1}{4}1/(z_1)^2F_{\mu\nu}'F^{\prime \mu\nu}+1/(z_2)^2\partial^{\mu}\Phi^{\prime*}\partial_{\mu}\Phi'-
\nonumber\\
&&-i e/(z_1z_2^2) A^{\prime\mu}\Phi^{\prime*}\partial_{\mu}\Phi'+ie/(z_1z_2^2)e A_{\mu}'\Phi'\partial^{\mu}\Phi^{\prime*}+e^2(A^{\prime\mu}A_{\mu}')^2,
\label{lagr434}
\end{eqnarray}
where $z_1\approx 1$ and $z_2\approx 0$. The kinetic term for the Higgs scalar is very big whereas that for the gauge field is very small. Alternatively  the interaction term $A_{\mu}A^{\mu}A_{\nu}A^{\nu}$ is very large. The Higgs expectation value is very big as it can be seen from Eq. (\ref{Higgs435}).  The structure of the Lagrangian and the large vev of the Higgs indicate that the system is in a combined Higgs confinement phase.

\subsection{Confinement phase}
The parameters $x$ and $y$ take the values: $y \approx 1/2$ and $y \approx 1/2$.
The Lagrangian for this case has the expression:
\begin{eqnarray}
&&{\cal L}=-\frac{1}{4}1/(z_1)^2F_{\mu\nu}'F^{\prime\mu\nu}+1/(z_2)^2\partial^{\mu}\Phi^{\prime*}\partial_{\mu}\Phi'-
\nonumber\\
&&-i e/(z_1z_2^2) A^{\prime\mu}\Phi^{\prime*}\partial_{\mu}\Phi'+ie/(z_1z_2^2)eA_{\mu}'\Phi'\partial^{\mu}\Phi^{\prime*}+\frac{1}{2}e^2A_{\mu}'A^{\prime\mu}\Phi^{\prime*}\Phi'-m^2/z_2^2\Phi^{\prime*}\Phi'
++e^2(A^{\prime\mu}A_{\mu}')^2
\label{lagr434}
\end{eqnarray}
where $z_1\approx \sqrt{1/2}$ and $z_2\approx \sqrt{1/2}$. In this case both the kinetic scalar term and the kinetic gauge terms are small. We first rescale the scalar field as $e\Phi\rightarrow\Phi$. Thus one can neglect all the terms containing the scalar field except that containing the quadrilinear interaction with the gauge field. The approximate Lagrangian in the old variables $A_{\mu}$ and $\Phi$ is:
\begin{eqnarray}
{\cal L}=\frac{1}{2}A^{\nu}\partial^2 A_{\nu}+\frac{e^2}{8}(A^{\mu}A_{\mu})^2+\frac{1}{2}A^{\mu}A_{\mu}|\Phi|^2,
\label{lagr1234}
\end{eqnarray}
where we considered the Feynman gauge. We shall use the same a approach as in Eqs. (\ref{rez324567}), (\ref{rez11213}) and (\ref{qw234})  to deal with the $A^{\mu}A_{\mu}|\Phi|^2$ term and with the integral over $\Phi(x)$ to obtain:
\begin{eqnarray}
Z=c+\int d A^{\mu}(x)\exp[i[\int d^4 x \frac{1}{2}A^{\nu}(x)\partial^2A_{\nu}(x)+\frac{e^2}{8}(A^{\mu}(x)A_{\mu}(x))^2],
\label{part4567}
\end{eqnarray}
where $c$ is a large constant.
We shall rewrite the second term on the right hand side of the Eq.(\ref{part4567}) as:
\begin{eqnarray}
\int d A^{\mu} \exp[i[\int d^4 xd^4y \frac{1}{2}A^{\nu}(x)\partial^2(x)\delta(x-y)A_{\nu}(y)+\frac{e^2}{8}\int d^4x \int d^4 y(A^{\mu}(x)A_{\mu}(y))^2\delta(x-y)]],
\label{part2345}
\end{eqnarray}
and define the operator $K(x,y)=\partial^2\delta(x-y)$. In order to determine the expression in Eq. (\ref{part2345}) it is easier to work in the coordinate space.
We write:
\begin{eqnarray}
&&\int d^4 x \int d^4 y\frac{1}{2}A^{\nu}(x)\partial^2(x)\delta(x-y)A_{\nu}(y)+\frac{e^2}{8}\int d^4x \int d^4 y(A^{\mu}(x)A_{\mu}(y))^2\delta(x-y)=
\nonumber\\
&&\int d^4 x \int d^4y \frac{e^2}{8}[A_{\mu}(x)A^{\mu}(y)+\frac{2}{e^2}\partial(y)\partial(x)]^2\delta(x-y)-\frac{1}{2e^2}\partial^2(x)\partial^2(y)\delta(x-y).
\label{id54678}
\end{eqnarray}
First we need to integrate:
\begin{eqnarray}
\int d A^{\mu}(x) \exp[i[\int d^4 x \int d^4y \frac{e^2}{8}[A_{\mu}(x)A^{\mu}(y)+\frac{2}{e^2}\partial(y)\partial(x)]^2\delta(x-y)]].
\label{newexpr768905}
\end{eqnarray}
For that we make the change of variable:
\begin{eqnarray}
A_{\mu}(x)\Rightarrow A_{\mu}(x)-\partial_{\mu}(x).
\label{ch5678}
\end{eqnarray}
This eliminates the dependence on the operator $\partial^2$ of the integral in Eq. (\ref{newexpr768905}) and leads to:
\begin{eqnarray}
&&\int d A^{\mu} \exp[i[\int d^4 x \int d^4y \frac{e^2}{8}[A_{\mu}(x)A^{\mu}(y)+\frac{2}{e^2}\partial(y)\partial(x)]^2\delta(x-y)]]=
\nonumber\\
&&\int d A^{\mu} \exp[i[\int d^4 x \int d^4y \frac{e^2}{8}\delta(x-y)[A_{\mu}(x)A^{\mu}(y)]^2]=d
\label{rez32456}
\end{eqnarray}
The Lorentz invariant two point function in the coordinate space is just:
\begin{eqnarray}
\langle A^{\mu}(x)A_{\mu}(z)\rangle={\rm const} \int d^4 y \frac{1}{Z}\frac{\delta(y-z)\delta Z}{\delta K(x,y)}\approx d\frac{-i}{2e^2}K(x,z)/(c+d)={\rm const} K(x,z)
\label{rez4567}
\end{eqnarray}
 Moreover knowing that the inverse of the operator K is the propagator in the functional approach,
\begin{eqnarray}
&&\int d^4 z K(x,z)\Delta(z,y)=\delta^4(x-y)
\nonumber\\
&&\Delta(x,y)=\int \frac{d^4 k}{(2\pi)^4}e^{ik(x-y)}\frac{1}{k^2-\mu^2}
\label{rez54678}
\end{eqnarray}
where $\mu^2$ is the infrared regulator, we get:
  \begin{eqnarray}
 \langle A^{\mu}(x)A_{\mu}(y)\rangle\approx  K(x,y)\approx \frac{1}{\Delta(x,y)}
 \label{finalres54678}
 \end{eqnarray}
We notice that in the coordinate space we obtain an expression for the two point function that is the inverse of the regular propagator.  Then can write directly that in spherical coordinates (we consider as sources two scalar fields with the momenta $q_1$, $q_2$ such that $p^2=(q_1-q_2)^2\approx |\vec{q_1}-\vec{q_2}|^2=|\vec{p}|^2$) the corresponding potential in first order is :
\begin{eqnarray}
V(r)\approx r
\label{pot7689}
\end{eqnarray}
This can also be observed directly form the fact that for a massless particle the usual first order propagator behaves like $\frac{1}{r}$ ( and in our notation is like $\frac{1}{K(x,y)}$ so  the inverse of this (our result in Eq.(\ref{rez4567})) behaves like $r$.

\section{Conclusions}

The phase structure of the abelian Higgs model is mostly known on a lattice, for fixed  radial component of the scalar field or through numerical simulations. In the present work we do not employ any of these artifices to study the behavior of the system. Instead by using the dual description of the scalar and gauge fields in terms of fermion variables we extend  the initial Lagrangian to take into account the various terms that might appear through a Fierz rearrangement of the fields. The most general Lagrangian obtained in this way is no longer gauge invariant but neither are all the phases in which the model is in.

By tuning the contributions of the interaction terms that are in the Lagrangian the system passes through three main phases which coincide exactly to those described in the literature: Higgs, confinement and Coulomb. However each phase has its own approximate Lagrangian which differs from one phase to another allowing one to determine the order parameter $\langle A^{\mu}(x)A_{\mu}(y)\rangle$ in the standard functional approach. It turns out that the corresponding potentials have the well know behavior for each phase. There is also a fourth region in the parameter space where both the Higgs and confinement phases coexist showing that there is no clear distinction between the two of them.

What is particular to our approach besides the overall treatment is that the $\lambda$ term in the Higgs potential plays no role whatsoever. The initial Lagrangian does not display spontaneous symmetry breaking (has the wrong mass term) but its overall modified structure leads to an actual Higgs phase.

The method is easy applicable to the pure abelian gauge model, electrodynamics without matter and may be extended to non abelian gauge invariant Lagrangians.

\section*{Acknowledgement}

The work of R. J. was supported by a grant of the Ministry of National Education, CNCS-UEFISCDI, project number PN-II-ID-PCE-2012-4-0078.

\end{document}